\documentclass[preprint2]{aastex}
\usepackage{lipsum}
\usepackage{graphicx}
\usepackage{pfnote}

\shorttitle{Gamma-ray observations of Tycho's SNR with VERITAS and Fermi}
\shortauthors{The VERITAS Collaboration}

\begin{document}


\title{Gamma-ray observations of Tycho's supernova remnant with VERITAS and Fermi}

\author{
S.~Archambault\altaffilmark{1},
A.~Archer\altaffilmark{2},
W.~Benbow\altaffilmark{3},
R.~Bird\altaffilmark{4},
E.~Bourbeau\altaffilmark{1},
M.~Buchovecky\altaffilmark{4},
J.~H.~Buckley\altaffilmark{2},
V.~Bugaev\altaffilmark{2},
M.~Cerruti\altaffilmark{3},
M.~P.~Connolly\altaffilmark{5},
W.~Cui\altaffilmark{6,7},
V.~V.~Dwarkadas\altaffilmark{8},
M.~Errando\altaffilmark{2},
A.~Falcone\altaffilmark{9},
Q.~Feng\altaffilmark{1},
J.~P.~Finley\altaffilmark{6},
H.~Fleischhack\altaffilmark{10},
L.~Fortson\altaffilmark{11},
A.~Furniss\altaffilmark{12},
S.~Griffin\altaffilmark{1},
M.~H{\"u}tten\altaffilmark{10},
D.~Hanna\altaffilmark{1},
J.~Holder\altaffilmark{13},
C.~A.~Johnson\altaffilmark{14},
P.~Kaaret\altaffilmark{15},
P.~Kar\altaffilmark{16},
N.~Kelley-Hoskins\altaffilmark{10},
M.~Kertzman\altaffilmark{17},
D.~Kieda\altaffilmark{16},
M.~Krause\altaffilmark{10},
S.~Kumar\altaffilmark{13},
M.~J.~Lang\altaffilmark{5},
G.~Maier\altaffilmark{10},
S.~McArthur\altaffilmark{6},
A.~McCann\altaffilmark{1},
P.~Moriarty\altaffilmark{5},
R.~Mukherjee\altaffilmark{18},
D.~Nieto\altaffilmark{19},
S.~O'Brien\altaffilmark{16},
R.~A.~Ong\altaffilmark{4},
A.~N.~Otte\altaffilmark{21},
N.~Park\altaffilmark{22}\altaffilmark{*},
M.~Pohl\altaffilmark{23,10},
A.~Popkow\altaffilmark{4},
E.~Pueschel\altaffilmark{20},
J.~Quinn\altaffilmark{20},
K.~Ragan\altaffilmark{1},
P.~T.~Reynolds\altaffilmark{24},
G.~T.~Richards\altaffilmark{21},
E.~Roache\altaffilmark{3},
I.~Sadeh\altaffilmark{10},
M.~Santander\altaffilmark{18},
G.~H.~Sembroski\altaffilmark{6},
K.~Shahinyan\altaffilmark{11},
P.~Slane\altaffilmark{25},
D.~Staszak\altaffilmark{22},
I.~Telezhinsky\altaffilmark{23,10},
S.~Trepanier\altaffilmark{1},
J.~Tyler\altaffilmark{1},
S.~P.~Wakely\altaffilmark{22},
A.~Weinstein\altaffilmark{26},
T.~Weisgarber\altaffilmark{27},
P.~Wilcox\altaffilmark{15},
A.~Wilhelm\altaffilmark{23,10},
D.~A.~Williams\altaffilmark{14},
B.~Zitzer\altaffilmark{1}
}

\altaffiltext{1}{Physics Department, McGill University, Montreal, QC H3A 2T8, Canada}
\altaffiltext{2}{Department of Physics, Washington University, St. Louis, MO 63130, USA}
\altaffiltext{3}{Fred Lawrence Whipple Observatory, Harvard-Smithsonian Center for Astrophysics, Amado, AZ 85645, USA}
\altaffiltext{4}{Department of Physics and Astronomy, University of California, Los Angeles, CA 90095, USA}
\altaffiltext{5}{School of Physics, National University of Ireland Galway, University Road, Galway, Ireland}
\altaffiltext{6}{Department of Physics and Astronomy, Purdue University, West Lafayette, IN 47907, USA}
\altaffiltext{7}{Department of Physics and Center for Astrophysics, Tsinghua University, Beijing 100084, China.}
\altaffiltext{8}{Department of Astronomy and Astrophysics, University of Chicago, Chicago, IL, 60637}
\altaffiltext{9}{Department of Astronomy and Astrophysics, 525 Davey Lab, Pennsylvania State University, University Park, PA 16802, USA}
\altaffiltext{10}{DESY, Platanenallee 6, 15738 Zeuthen, Germany}
\altaffiltext{11}{School of Physics and Astronomy, University of Minnesota, Minneapolis, MN 55455, USA}
\altaffiltext{12}{Department of Physics, California State University - East Bay, Hayward, CA 94542, USA}
\altaffiltext{13}{Department of Physics and Astronomy and the Bartol Research Institute, University of Delaware, Newark, DE 19716, USA}
\altaffiltext{14}{Santa Cruz Institute for Particle Physics and Department of Physics, University of California, Santa Cruz, CA 95064, USA}
\altaffiltext{15}{Department of Physics and Astronomy, University of Iowa, Van Allen Hall, Iowa City, IA 52242, USA}
\altaffiltext{16}{Department of Physics and Astronomy, University of Utah, Salt Lake City, UT 84112, USA}
\altaffiltext{17}{Department of Physics and Astronomy, DePauw University, Greencastle, IN 46135-0037, USA}
\altaffiltext{18}{Department of Physics and Astronomy, Barnard College, Columbia University, NY 10027, USA}
\altaffiltext{19}{Physics Department, Columbia University, New York, NY 10027, USA}
\altaffiltext{20}{School of Physics, University College Dublin, Belfield, Dublin 4, Ireland}
\altaffiltext{21}{School of Physics and Center for Relativistic Astrophysics, Georgia Institute of Technology, 837 State Street NW, Atlanta, GA 30332-0430}
\altaffiltext{22}{Enrico Fermi Institute, University of Chicago, Chicago, IL 60637, USA}
\altaffiltext{23}{Institute of Physics and Astronomy, University of Potsdam, 14476 Potsdam-Golm, Germany}
\altaffiltext{24}{Department of Physical Sciences, Cork Institute of Technology, Bishopstown, Cork, Ireland}
\altaffiltext{25}{Harvard-Smithsonian Center for Astrophysics, 60 Garden Street, Cambridge, MA 02138, USA}
\altaffiltext{26}{Department of Physics and Astronomy, Iowa State University, Ames, IA 50011, USA}
\altaffiltext{27}{Department of Physics, University of Wisconsin-Madison, Madison, WI 53706, USA}

\altaffiltext{*}{E-mail: nahee@uchicago.edu}

\setcounter{footnote}{0}

\begin{abstract}
High-energy gamma-ray emission from supernova remnants (SNRs) has provided a unique perspective for studies of Galactic cosmic-ray acceleration. Tycho's SNR is a particularly good target because it is a young, type Ia SNR that is well-studied over a wide range of energies and located in a relatively clean environment. Since the detection of gamma-ray emission from Tycho's SNR by VERITAS and \textit{Fermi}-LAT, there have been several theoretical models proposed to explain its broadband emission and high-energy morphology. We report on an update to the gamma-ray measurements of Tycho's SNR with 147 hours of VERITAS and 84 months of \textit{Fermi}-LAT observations, which represents about a factor of two increase in exposure over previously published data. About half of the VERITAS data benefited from a camera upgrade, which has made it possible to extend the TeV measurements toward lower energies. The TeV spectral index measured by VERITAS is consistent with previous results, but the expanded energy range softens a straight power-law fit. At energies higher than 400 GeV, the power-law index is $2.92 \pm 0.42_{\mathrm{stat}} \pm 0.20_{\mathrm{sys}}$. 
It is also softer than the spectral index in the GeV energy range, $2.14 \pm 0.09_{\mathrm{stat}} \pm 0.02_{\mathrm{sys}}$, measured by this study using \textit{Fermi}--LAT data. The centroid position of the gamma-ray emission is coincident with the center of the remnant, as well as with the centroid measurement of \textit{Fermi}--LAT above 1 GeV. The results are consistent with an SNR shell origin of the emission, as many models assume. 
The updated spectrum points to a lower maximum particle energy than has been suggested previously. 
\end{abstract}


\keywords{supernova remnant: general -- supernova remnant: individual(Tycho's SNR) -- gamma rays: observations }

\section{Tycho's SNR as a Cosmic Ray accelerator}
Supernova remnants (SNRs) have been suggested to be the main accelerators of Galactic cosmic rays (CRs)~\citep{Ginzburg:1961bh}. Several studies have proposed that SNRs can efficiently convert the kinetic energy of the supernova explosion to generate relativistic CRs via diffusive shock acceleration~\citep{Bell:1978is, 1978MNRAS.182..443B, Schure:2012hj, Reynolds:2008fi}. 
Indirect evidence of the acceleration of the leptonic component of CRs up to 100 TeV has been provided by the detection of non-thermal X-ray emission from the rims of several remnants (e.g. ~\citealp{1995Natur.378..255K}). Corresponding evidence for hadronic acceleration has been elusive, but the improved sensitivity of GeV--TeV gamma-ray telescopes over the past decade has opened a new window to study the interactions of high-energy particles around SNRs. 

Gamma rays can be generated as bremsstrahlung radiation when electrons and positrons interact with ambient matter, or as a result of inverse Compton scattering of low energy photons around SNRs. Hadronic particle interactions can also create gamma rays via the pion-decay process. 
By combining our knowledge of the SNR environment with gamma-ray observations, we can study the acceleration and propagation of particles in and around the remnant.
The recent detection of a pion-decay signature from two middle-aged SNRs, IC 443 and W44, by \textit{Fermi}-LAT has demonstrated the existence of hadronic particle acceleration in SNRs~\citep{Ackermann:2013jq}. However, several questions remain to be answered, such as the maximum energies to which particles can be accelerated in SNRs, the efficiency of the acceleration in the remnants, and the nature of the acceleration process. Resolving these questions is necessary to determine whether SNRs are indeed the main accelerators of Galactic cosmic rays up to the ``knee" region ($\sim$ 3 PeV). 

Gamma-ray observations of young SNRs (with ages less than a few thousand years) can provide valuable data to address these questions. 
Young SNRs are usually well studied over a wide energy range. Non-thermal X-ray emission is detected from many of these objects, providing data to investigate the acceleration processes of the electrons, and to gauge the strength of the magnetic fields~\citep{2012SSRv..166..231R}. The ages of these remnants are also better constrained than those of older remnants. In particular, the ages of ``historical" SNRs are well known. These are all important ingredients that allow the development of detailed emission models.
Furthermore, young SNRs can accelerate particles to higher energies than older remnants can~\citep{Berezhko:1997dz,2012arXiv1206.5018D,Bell:2014gk}. Thus they serve as better probes of the maximum energy to which particles can be accelerated in SNRs.  

SNR G120.1+1.4, also known as ``Tycho's supernova remnant" (hereafter referred to as Tycho), is one of the best-studied young SNRs. It is the remnant of a historical supernova that was observed in 1572. The historical light curve records~\citep{Baade:1945gn} and ejecta composition measurements in the X-ray band~\citep{Decourchelle:2001if} suggested a Type Ia origin, which was confirmed by spectroscopic analysis of the light echo from the explosion~\citep{Krause:2008fr}. 
The radio and X-ray expansion rate measurements suggest that the global evolutionary state of Tycho is pre-Sedov, while local regions with higher density are evolutionarily more advanced~\citep{Aharonian:2001kq}.

X-ray images for energies higher than 4 keV measured by \textit{Chandra} show thin filamentary structures in the rim of Tycho~\citep{2002ApJ...581.1101H}, which have been interpreted as non-thermal X-ray emission generated by high-energy electrons~\citep{2005ApJ...621..793B}. \cite{Parizot:2006dz} estimated the magnetic field strength at the rim of the SNR to be about 200 $\mu$G, assuming that the widths of the filament structures are due to radiative energy loss of high-energy electrons. 
For such a strong magnetic field, the radiative losses limit the maximum electron energy. As a result, the maximum energies of electrons and protons can be different. 
\cite{Parizot:2006dz}  also estimated the maximum electron energy (hereafter defined as the cut-off energy for primary particles following a power-law distribution with an exponential cut-off) to be 5--7 TeV by using the X-ray cut-off energy obtained by comparison of X-ray fluxes and radio fluxes. Based on their estimation of the magnetic field strength, diffusion coefficient, and X-ray spectral cut-off energy, the maximum energy of accelerated protons in the remnant was estimated to be in the range of 100 TeV--2 PeV. 
An alternative explanation for such thin filaments was given by \cite{2005ApJ...626L.101P}, who suggested that they may be the result of magnetic field damping, in which case the magnetic field may not be as high and the acceleration of particles would be less efficient. 
\cite{2007ApJ...665..315C} studied the intensity profile of radio and X-ray bands at the rim with a hydrodynamic model to test these two scenarios, suggesting a combination of cooling and rapid damping to explain the filament structures. 

Deep observations with \textit{Chandra} have revealed regular patterns, or ``stripes", of non-thermal emission. \cite{2011ApJ...728L..28E} interpreted the gaps between these stripes as arising from the gyration of high-energy protons in a magnetic field, providing evidence of proton acceleration up to 100 TeV--1 PeV. In contrast, \cite{Bykov:2011jl} explained these stripes as the result of magnetic field turbulence. Although their explanation is different, they estimated the maximum proton energy responsible for the stripes to be also on the order of 1 PeV. 

Recent \textit{NuSTAR} observations have been used to study the correlation of shock velocity and expansion parameters with measurements of the X-ray spectral rolloff energy~\citep{2015ApJ...814..132L}. The rolloff energy $E_\textnormal{rolloff}$ is a characteristic synchrotron cut-off energy proportional to $BE_\textnormal{max}^2$, where $B$ is the magnetic field strength and $E_\textnormal{max}$ is an assumed exponential cut-off energy in the electron spectrum~\citep{Reynolds:1999ef}. The authors suggested that the scenario of the maximum electron energy being limited by the age of SNR, instead of by radiative energy loss of electrons, best fits the data.  They estimated the maximum energy of electrons (and protons, in this case) to be 5--12 TeV. This led to an estimation of the magnetic field strength of around 30 $\mu$G, which is lower than suggested by a radio and X-ray morphology study which included both energy loss and magnetic damping scenarios~\citep{2015ApJ...812..101T}. Also, it is lower than the minimum magnetic field strength of 80~$\mu$G required to explain the multi-wavelength emission from radio to TeV gamma-ray energies, as suggested by \cite{Acciari:2011iq}. \cite{2015ApJ...814..132L} noted that, alternatively, a higher magnetic field and a loss-limited maximum electron energy can be accommodated if the correlation of X-ray rolloff energy with the shock velocity arose from the obliquity effect---the acceleration rate dependency on the angle between the shock front and the local magnetic field.

Detections of gamma-ray emission at TeV~\citep{Acciari:2011iq} and GeV~\citep{2012ApJ...744L...2G} energies provide additional data to study the acceleration of high-energy particles in Tycho. The gamma-ray measurements also provide another diagnostic by which to estimate the maximum energy of hadrons in the remnant. 

Several models were developed to explain the gamma-ray emission from Tycho~\citep{2014ApJ...783...33S,2013ApJ...763...14B, 2013MNRAS.429L..25Z,Morlino:2012km,Atoyan:2012bz}, including two detailed studies~\citep{Morlino:2012km,2014ApJ...783...33S} that modeled the full spectral energy distribution of Tycho from radio to gamma-ray energies, along with the morphology of the radio and X-ray emission. 
Most of these models, with one exception~\citep{Atoyan:2012bz}, conclude that the gamma-ray emission from Tycho is predominantly produced by hadronic interactions, although the details of the models vary considerably. \cite{Morlino:2012km} estimated the maximum proton energy to be 470 TeV based on a semi-analytical calculation,  while \cite{2014ApJ...783...33S} estimated it to be 50 TeV from a full hydrodynamic simulation.  

The shape of Tycho in the radio band and in X-rays is roughly spherical. Detailed regional expansion rate differences measured in radio~\citep{Reynoso:1997bi} and X-ray~\citep{2010ApJ...709.1387K} suggest that the northern, northeastern, and eastern parts of the shell of the remnant may be expanding into denser regions compared to the southern parts of the remnant. A recent study by \cite{2013ApJ...770..129W} reported on the existence of an azimuthal density variation around the rim of Tycho using \textit{Spitzer} data, showing that the northeastern region has 3--10 times higher density compared to the southwestern region of the remnant.
 
A large molecular cloud was observed near Tycho in the north/northeastern region. Interactions between the northeastern region of the remnant and the molecular cloud were suggested based on radio HI and CO measurements~\citep{Reynoso:1999bw,Lee:2004gs}. However, these were not confirmed in later measurements~\citep{Tian:2011ez}.

The TeV gamma-ray image of Tycho presented in 2011 by VERITAS shows the morphology of the emission to be compatible with a point source. The peak of the TeV emission shows indications of being offset from the center of the remnant towards the northeastern part of the remnant, where the density of the surrounding medium is higher and the molecular cloud is observed along the line of sight. 
Although the measured offset, $0.04^\circ \pm 0.023^\circ_{\mathrm{stat}} \pm 0.014^\circ_{\mathrm{sys}}$, is not statistically different from zero, this introduced the question of whether the gamma-ray emission is dominated by pion decay resulting from hadronic interactions within the entire swept-up SNR shell, from the high density regions of the shell, or by interactions with the molecular cloud. 


VERITAS has conducted deeper observations of Tycho to improve both the flux and centroid measurements reported previously.
In this paper, we update our originally published high-energy gamma-ray results with a factor of two increase in VERITAS exposure, coupled with a \textit{Fermi}-LAT analysis with improved sensitivity and a deeper exposure than previously shown. We compare existing theoretical models with the updated measurements, study the maximum energy of particles that can be accelerated in Tycho, and discuss the origin of the gamma-ray emission.

\begin{figure}[t!]
    \centering
            \includegraphics[width=\linewidth]{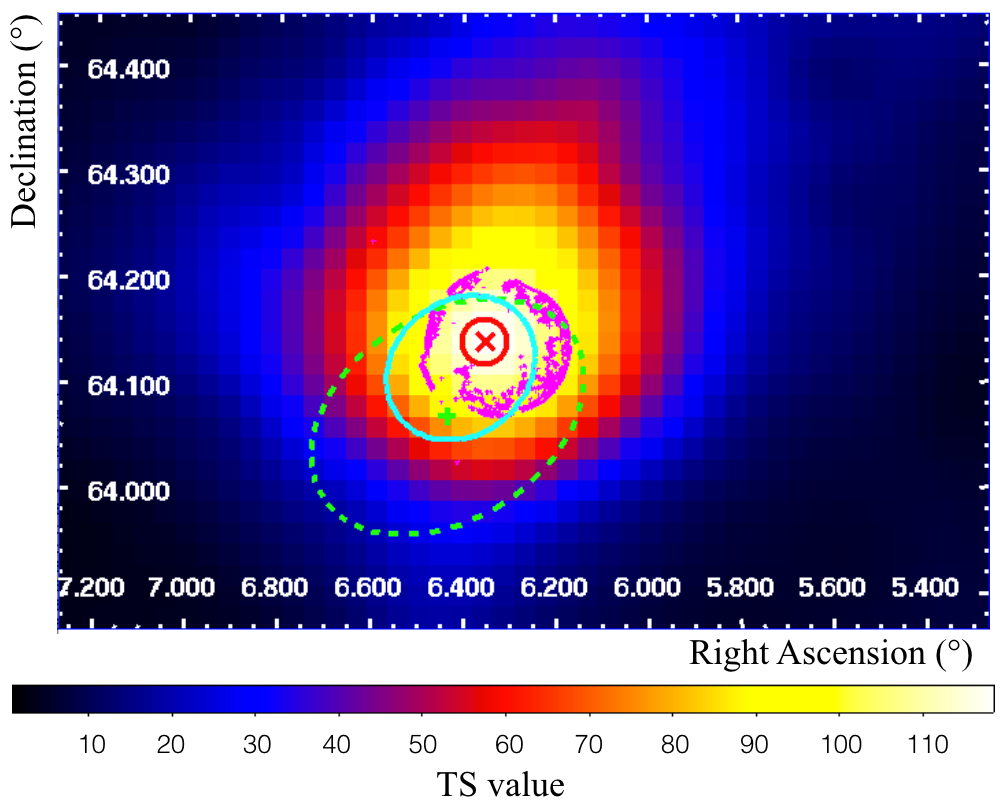}
        \caption{Smoothed \textit{Fermi} TS map with the P8R2$\_$CLEAN$\_$V6 IRF for energies higher than 1 GeV.  The map was smoothed with a Gaussian kernel with a radius of $0.06^\circ$. The magenta contours indicate the \textit{Chandra} X-ray intensity at energies above 4.1 keV\protect\footnote{\tiny{http://chandra.harvard.edu}}. The cyan line is the previously published $95\%$ confidence area for the \textit{Fermi}-LAT position~\citep{2012ApJ...744L...2G}. The centroid and error of 3FGL J0025.7+6404 are marked with a cross and dashed green line~\citep{Acero:2015ga}. The best-fit position and $68\%$ confidence level of this study are shown with a red cross mark and a red circle.}  
   \label{Fig:Tycho_skymap_Fermi}
 \end{figure}

\section{\textit{Fermi} observation of Tycho's SNR}
\subsection{Analysis}
The \textit{Fermi Gamma-ray Space Telescope} was launched in 2008 June. The principal \textit{Fermi} science instrument, the Large Area Telescope (\textit{Fermi}-LAT), has provided all-sky coverage in the 20 MeV to $>$300 GeV energy range over eight years. In June 2015, the \textit{Fermi} collaboration released ``Pass 8" LAT data for analysis~\citep{2013arXiv1303.3514A}. Pass 8 provides a larger effective area, especially for the lowest and highest energies, with an improved point spread function (PSF) compared to previous data releases~\citep{2013arXiv1303.3514A}. In this paper, we update our previous study~\citep{2015arXiv150807070P} to a Pass 8 analysis. 

We analyzed a dataset of 84 months, from 2008 August to 2015 August, selecting events with energies from 300 MeV to 500 GeV that fall within a radius of $25^\circ$ centered on the position of the remnant.  
The publicly available Fermi Science Tools\footnote{http://fermi.gsfc.nasa.gov/ssc} were used for the analysis. 

The recommended quality cuts for standard analysis were implemented. All sources from the 3FGL catalog~\citep{Acero:2015ga} that fall within a $40^\circ$ radius around Tycho were included in the analysis for modeling. A binned-likelihood analysis with a bin size of $0.1^\circ$ was performed first for the 3FGL source position associated with Tycho (3FGL J0025.7+6404). The results were later compared by repeating the analysis with the best-fit position of the emission from the Tycho region from this analysis of the \textit{Fermi}-LAT data.
The analysis method maximizes the likelihood of all sources within the region of interest for the given source model. Fluxes of the diffuse background emission as well as all sources except five weak sources located within a radius of $10^\circ$ around Tycho were allowed to vary for the likelihood analysis. Fluxes were also allowed to vary for sources located within a radius of $15^\circ$ around Tycho and that have a statistical significance higher than $15\sigma$ in the 3FGL catalog. The source spectral index was allowed to vary for sources located within a radius of $3^\circ$ around Tycho. These parameters were set to ensure the convergence of the likelihood fit. 

\begin{figure}[t!]
    \centering
            \includegraphics[width=\linewidth]{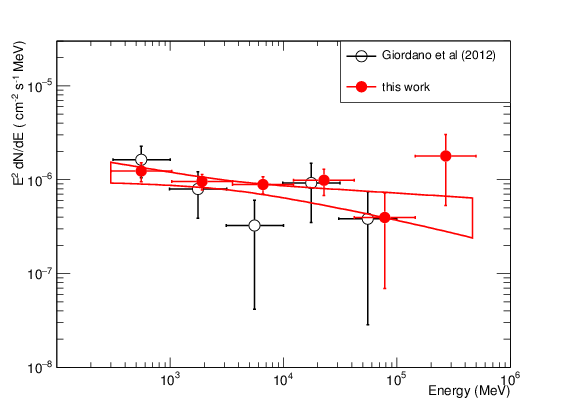}
        \caption{The updated Fermi spectrum overlaid with the previous results. The solid red line shows the $1\sigma$ statistical error band from the full fit.} 
   \label{Fig:Tycho_spectrum_Fermi}
 \end{figure}

\subsection{Results}
The statistical significance of the emission is calculated from the Test Statistic (TS) of the likelihood ratio test, defined as twice the difference of the log likelihood between the given source model and the null hypothesis. 
The overall fit from 300 MeV to 500 GeV yields a TS of 107 for 3FGL J0025.7+6404, assuming a power-law distribution of photon energies. The best source position was estimated with \textit{gtfindsrc} for energies higher than 1 GeV by using the P8R2$\_$CLEAN$\_$V6 instrument response function (IRF). This IRF was chosen because its smaller PSF can provide a sharper image than the standard IRF, P8R2$\_$SOURCE$\_$V6.  
The result showed the best-fit position of Right Ascension (RA) $0^h25^m24^s.39$ and declination $64^{\circ} 8{\arcmin} 25{\arcsec}$ with a $68\%$ statistical error of $0.02^\circ$. The TS value of the likelihood analysis at the best-fit position was 130.  

Figure~\ref{Fig:Tycho_skymap_Fermi} shows a TS map generated using the \textit{Fermi} tool \textit{gttsmap} for the region around Tycho.
The TS maps for two different IRFs, P8R2$\_$CLEAN$\_$V6 and P8R2$\_$SOURCE$\_$V6, were checked for energies higher than 1 GeV, and both agree within the statistical errors. 
The $68\%$ confidence region from this study agrees with the results of the previous paper, as well as with the position of the 3FGL source associated with Tycho. The centroid position coincides with the centroid of the X-ray emission within the statistical error. 

While we used the P8R2$\_$CLEAN$\_$V6 IRF for the source position study, we report spectral results from the best-fit position analyzed with the P8R2$\_$SOURCE$\_$V6 IRF to increase the number of events. For the spectral studies, systematic uncertainties were estimated as the root-mean-square (RMS) of results from six different analyses that used different IRFs, two different regions of interest, and the maximum and minimum of the systematic error on the effective area\footnote{\url{http://fermi.gsfc.nasa.gov/ssc/data/analysis/scitools/Aeff_Systematics.html}}.  The measured integral flux for energies higher than 300 MeV is ($3.60\pm0.62_{\mathrm{stat}}\pm0.12_{\mathrm{sys}})\times 10^{-9}$~$\mathrm{cm}^{-2}$~$\mathrm{s}^{-1}$, assuming a power-law distribution of photon energies. The estimated spectral index is $2.14\pm0.09_{\mathrm{stat}}\pm0.02_{\mathrm{sys}}$. The results agree within 1$\sigma$ with the discovery paper~\citep{2012ApJ...744L...2G} and with previous results with Pass 7 reprocessed data~\citep{2015arXiv150807068P}. 

The entire energy range was  divided into evenly spaced energy bins (in logarithmic scale) to compute the spectral energy distribution (SED). An individual likelihood analysis was performed for each bin using the fitted spectral parameters from the analysis of the entire energy range. All parameters except the flux of Tycho were fixed. The flux was calculated for bins with TS values higher than 4. Figure~\ref{Fig:Tycho_spectrum_Fermi} shows the SED with the P8R2$\_$SOURCE$\_$V6 IRF.  
Only statistical errors are shown in the figure since these dominate over the systematic uncertainties considered in this study. 

\section{VERITAS observation of Tycho's SNR}
VERITAS is an array of four atmospheric Cherenkov telescopes located at the Fred Lawrence Whipple Observatory in southern Arizona~\citep{Weekes:2002hw}. The telescope is designed to study astrophysical sources of gamma-ray emission in the 85 GeV--30 TeV range by detecting the Cherenkov light generated by air showers, cascades resulting from the interactions of the gamma rays in the atmosphere. Each of the four telescopes covers a field of view of $3.5^{\circ}$ with a  499-pixel photomultiplier tube (PMT) camera at the focal plane, collecting light from a 12 meter diameter reflector consisting of segmented mirrors. A coincident Cherenkov signal that triggers at least two out of four telescopes is required to trigger an array-wide read-out of the PMT signals~\citep{Holder:2006uh}. In its current configuration, the array has the sensitivity to detect a point source with a flux of $1\%$ of the Crab Nebula flux within 25 hours, and has an angular resolution better than $0.1^\circ$ at 1 TeV. 

The current performance of VERITAS is a result of two major hardware upgrades that changed the properties of the telescope array significantly: the relocation of one of the telescopes in 2009~\citep{2009arXiv0912.3841P} and a camera upgrade in 2012~\citep{2013arXiv1308.4849D}. 
The telescope relocation improved the overall angular resolution and the selection efficiency for gamma rays by making the array more symmetric. This enhanced the ability to measure morphological information of the gamma-ray showers. The VERITAS cameras  were equipped with high quantum efficiency PMTs during the 2012 upgrade. This made the array more sensitive to weaker signals, thus lowering the energy threshold of the array and improving the background rejection power in the low energy range~\citep{2015arXiv150807070P}. 

\subsection{Observation}
VERITAS has observed Tycho since 2008, collecting a total of 147 hours of data over five observing seasons, spanning both major upgrades of the array. The discovery of gamma-ray emission from Tycho was reported by VERITAS based on 67 hours of observation during 2008-2010. VERITAS has accumulated a total of 80 more hours since the detection paper, 74 hours of which were collected following the 2012 upgrade, with enhanced sensitivity at energies lower than a few TeV. 

Data were collected as close as possible to Tycho's culmination, resulting in an average elevation of $55^{\circ}$ over all observations. Observations were performed in ``wobble" mode, in which the telescope is pointed $0.5^{\circ}$ away from the target in the four cardinal directions~\citep{1994APh.....2..137F}.

\subsection{Analysis}
A standard Hillas moment analysis has been used for this study~\citep{1985ICRC....3..445H}. 
A detailed description of the VERITAS data analysis procedure can be found in \cite{2008ICRC....3.1325D}, and a description of the analysis tools can be found in \cite{Cogan:2007tk}. 
Cuts for the analysis were selected \textit{a priori} to provide good sensitivity for a point source with $0.9\%$ of the gamma-ray flux of the Crab Nebula.  Cuts were optimized, using Crab Nebula data, separately for the 2009--2011 dataset and for the 2012--2015 dataset to account for instrumental changes due to the hardware upgrade. The optimized cuts for 2009--2011 were also used for the 2008--2009 data after verifying their sensitivity on Crab Nebula data from this earlier period. Cuts were optimized to achieve a good compromise between broadband differential sensitivity and a low threshold energy. As a result, cuts for the 2009--2011 dataset have an energy threshold value at an elevation angle corresponding to the observations of Tycho of 800 GeV, similar to the analysis presented in the discovery paper, while cuts for the 2012--2015 set have a lower energy threshold value of 400 GeV with similar sensitivity. A cut on the angular distance from the test position to the reconstructed arrival direction of the shower was set to be 0.1$^{\circ}$ for both sets of cuts. Results were verified with an independent secondary analysis~\citep{2016ED_VERITAS}.

\begin{figure}[t!]
    \centering
            \includegraphics[width=\linewidth]{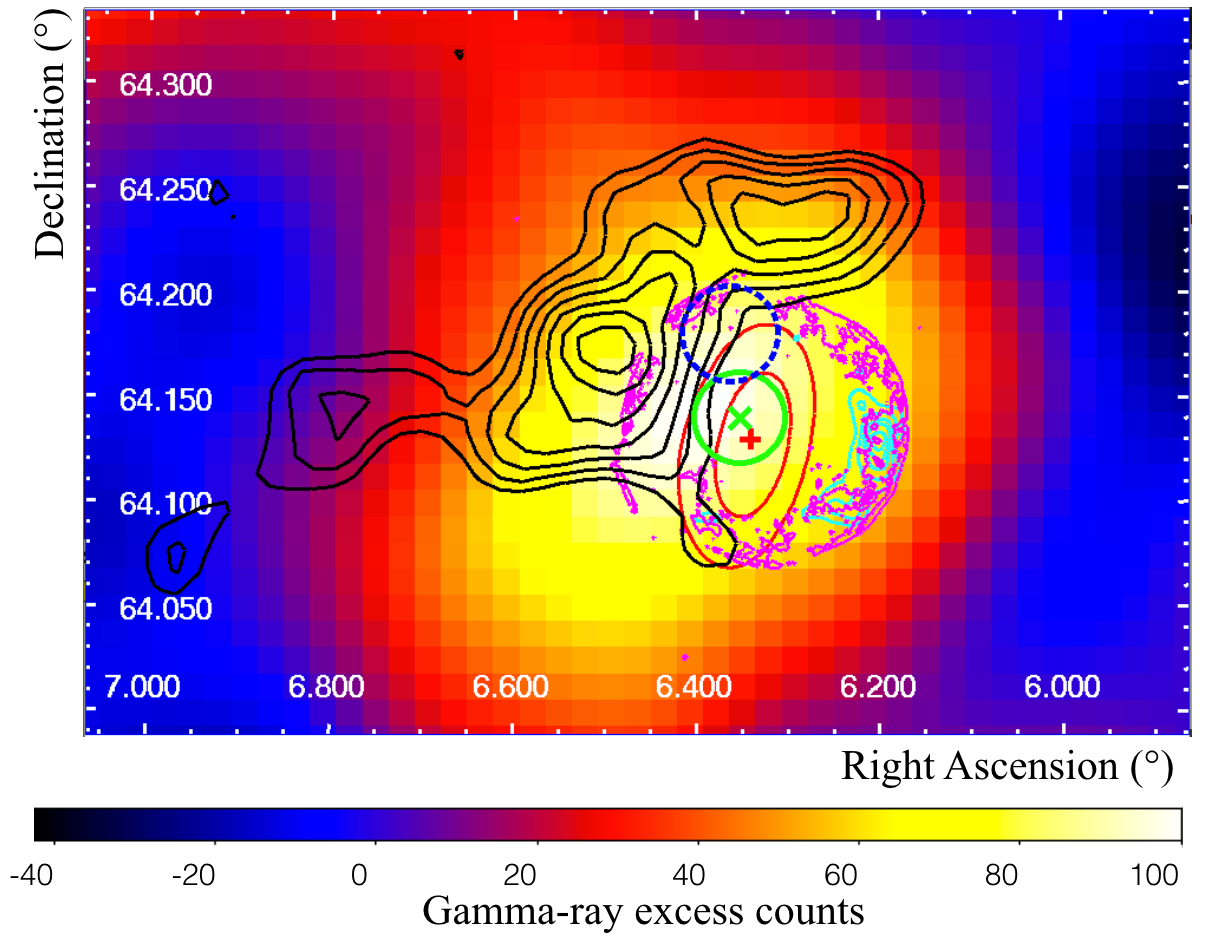}
        \caption{Smoothed VERITAS gamma-ray count map of the region around Tycho's SNR. The $1\sigma$ statistical error on the centroid position obtained by \cite{Acciari:2011iq} is drawn with a blue dashed circle. The updated centroid position is marked with a red cross and 68 and 95\% confidence levels of the position are shown with red contours. Each contour was determined from a fit with two degrees of freedom. \textit{Chandra}'s measurement of the X-ray emission with energies larger than 4.1 keV is shown by the magenta contours. Black contours are the $^{12}CO$ (J=1-0) emission integrated over the velocity range -68 km s$^{-1}$ to -50 km s$^{-1}$ using the measurements from the Five College Radio Astronomy Observatory Survey~\citep{Heyer:1998ih}. \textit{NuSTAR}'s measurements of X-rays in the energy range between 20 keV and 40 keV~\citep{2015ApJ...814..132L} after smoothing are shown by the cyan contours. The best-fit position and $68\%$ confidence level of the updated \textit{Fermi} analysis are shown as a green x mark and circle. }
        \label{Fig:Tycho_skymap_VERITAS}
 \end{figure}

\subsection{Results}
The analysis of the combined set of VERITAS data detected gamma-ray emission from Tycho with a significance of $6.9\sigma$. Figure~\ref{Fig:Tycho_skymap_VERITAS} shows the gamma-ray count map with the previously published centroid position and the updated centroid position. The map was smoothed with a Gaussian kernel with a radius of $0.06^{\circ}$.

The centroid position is estimated by maximizing the likelihood value of the data for a given background model and a source model. The background model was constructed from the data by estimating the spatial distribution of events outside of the source region, which was defined as a circle with a radius of $0.3^\circ$ around the center of Tycho.
For the source model, it is assumed that the gamma-ray distribution is produced by an unresolved point source. 
In this case, the source model can be described as an instrumental PSF. The PSF is described by a two-dimensional King function,  

\begin{displaymath} 
K(r) = N_{\mathrm{0}}(1+ (r/r_{\mathrm{0}})^{2})^{-\beta}
\end{displaymath} 
where $N_{\mathrm{0}}$ is a normalization factor, $r$ is an angular distance from the centroid, $r_{\mathrm{0}}$ is a radius, and $\beta$ is an index.

Two parameters, the radius and index, which define the shape of the PSF, are fixed to the best-fit values from a fit to simulated data. The simulated data were weighted to match the observational elevation and azimuth and measured spectral index of Tycho. 
This method assumes that measured event counts follow a Poisson distribution instead of a normal distribution, providing more accurate estimations of centroid positions compared to the method used for the previous paper~\citep{Acciari:2011iq}.

The centroid position reported was estimated only with the 2012--2015 dataset because it has the highest statistics and the best angular resolution.  
The estimated centroid is RA $0^h25^m21^s.60 \pm 7^s.20_{\mathrm{stat}}$ and declination $64^{\circ}7{\arcmin}48{\arcsec}  \pm 1{\arcmin}12{\arcsec}_{\mathrm{stat}}$. Statistical error contours of the 68\% and 95\% confidence levels are shown in Figure~\ref{Fig:Tycho_skymap_VERITAS} with the centroid. The updated centroid matches well with the center of the remnant. The uncertainty of VERITAS telescope pointing is $0.007^{\circ}$ as measured with stars after optical pointing offset correction~\citep{2015PhDT.......126G}. We estimate $0.006^{\circ}$ of combined systematic uncertainty on the centroid position from the shower reconstruction method and from the influence of the bin size of the count map used for the study. 

\begin{figure}[t!]
    \centering
            \includegraphics[width=\linewidth]{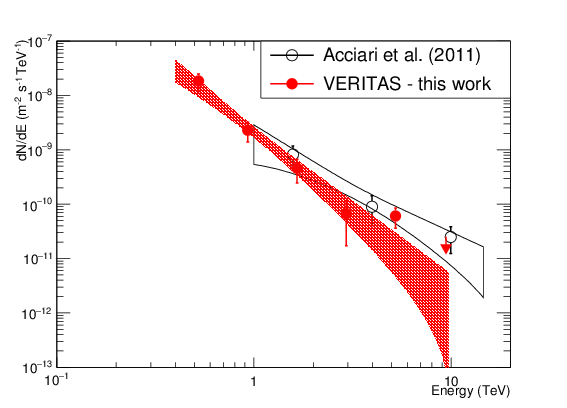}
        \caption{VERITAS spectra. The previous result is plotted as black empty circles and the result of the present study is shown with filled red circles. Flux errors were calculated from error propagation of the fitting function and drawn as a $1\sigma$ error band around the data points.}
        \label{Fig:Tycho_Spectrum_VERITAS}
 \end{figure}

The updated spectrum using all data is consistent with a power-law $dN/dE = N_{0}(E/1\ \mathrm{TeV})^{-\Gamma}$ with a normalization factor 
$N_0=(2.2\pm0.5_\mathrm{stat}\pm0.6_\mathrm{sys})\times10^{-13}$~$\mathrm{cm}^{-2}$~$\mathrm{s}^{-1}$~$\mathrm{TeV}^{-1}$ and a spectral index $\Gamma=2.92\pm0.42_\mathrm{stat}\pm 0.20_\mathrm{sys}$.
The reduced chi-square of the fit is 1.34 (4.01/3). Above 7.5 TeV, the gamma-ray excess has a significance of $\sim1\sigma$, and a $99\%$ confidence level upper limit of $2.5 \times 10^{-15}$~$\mathrm{cm}^{-2}$~$\mathrm{s}^{-1}$~$\mathrm{TeV}^{-1}$ was obtained by Rolke's method~\citep{Rolke:2005el} calculated with an index of $2.9$. The reduced significance of the data point at 10 TeV compared to the previous result is likely due to a statistical fluctuation. 
Figure~\ref{Fig:Tycho_Spectrum_VERITAS} shows the spectral analysis from this study in comparison with the previous result. 

Previous results reported a spectrum consistent with a power-law distribution with a spectral index of $1.95 \pm 0.51_{\mathrm{stat}} \pm 0.30_{\mathrm{sys}}$ for energies higher than 1 TeV. The updated result extends the measurement to lower energies, which was enabled by the camera upgrades of the VERITAS telescope. Flux measurements of the wider energy range extending from 400 GeV to 10 TeV reveal a softer index than previously reported. 

\section{Discussion}
\begin{figure}[t!]
    \centering
            \includegraphics[width=\linewidth]{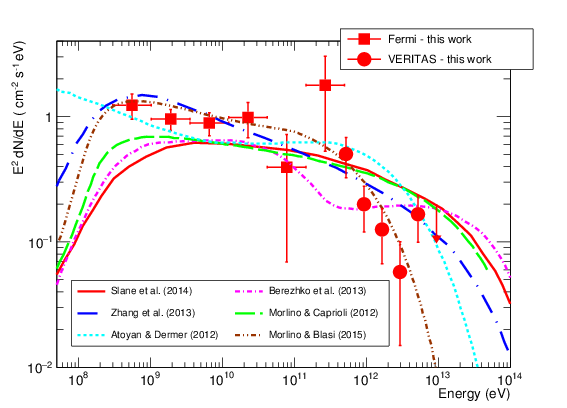}
        \caption{Fermi and VERITAS SEDs with theoretical models. Filled red squares show the \textit{Fermi} results and filled red circles show the VERITAS results from this study. The models discussed in the text appear as the solid red line (prefered model A from \cite{2014ApJ...783...33S}), the magenta short broken dashed line \citep{2013ApJ...763...14B}, the blue large broken dashed line \citep{2013MNRAS.429L..25Z}, the green dashed line \citep{Morlino:2012km}, the cyan dotted line (the leptonic model from \cite{Atoyan:2012bz}), and the brown double-broken dashed line (\cite{Morlino:2016kj} with a neutral fraction of 0.6).}
        \label{Fig:Tycho_CombinedSpectrum_wOldModels}
 \end{figure}

Figure~\ref{Fig:Tycho_CombinedSpectrum_wOldModels} shows the updated gamma-ray SEDs overlaid with the existing theoretical models. 
\cite{Morlino:2012km} took a semi-analytical approach to explain the morphology and flux of the multi-wavelength spectrum of Tycho from radio up to TeV energies, assuming that Tycho exploded in a homogeneous circumstellar medium. 
They postulated a distribution of high-velocity scattering centers throughout the cosmic-ray precursor and no motion in the downstream region, leading to a significant reduction in the compression ratio experienced by energetic particles and consequently to a soft power-law spectrum with a spectral index of 2.2.
\cite{2013ApJ...763...14B} explained the GeV--TeV flux by hadronic emission from a two-component medium, comprising a warm diffusive ISM and cold dense cloud clumps. 
Gamma-ray emission from these two media with different densities was used to obtain a gamma-ray spectral index of 2.0 in both the GeV and TeV range.
\cite{2013MNRAS.429L..25Z} suggested that the gamma-ray emission arises from cosmic-ray interactions with a cloud of density of 4--12 cm$^{-3}$, with an explosion energy conversion efficiency of 1\%.
While these models explained the gamma-ray emission via hadronic processes, \cite{Atoyan:2012bz} attempted to explain it with a pure leptonic model by introducing two emission zones with different properties. They argued that a realistic description of the non-thermal emission from a remnant with a spatially non-uniform magnetic field should at least consider two different emission zones with different magnetic fields and densities. 
\cite{2014ApJ...783...33S} provided the most detailed study, with a full hydrodynamic simulation including non-linear diffusive shock acceleration, to estimate both the thermal and non-thermal emission components from Tycho. Their simulation allowed electrons and hadrons to radiate in different environments and to be shocked at different times. Their best-fit model suggested that the GeV--TeV gamma-ray emission is dominated by a hadronic component.

All of the models described above were developed to explain the previously published GeV--TeV gamma-ray emission. The updated fluxes of TeV gamma-ray emission found in this paper for energies higher than 400 GeV are inconsistent with all these models. The models may need to be re-calculated to fit the updated gamma-ray spectra. 

\begin{figure} [t!] 
    \centering
            \includegraphics[width=\linewidth]{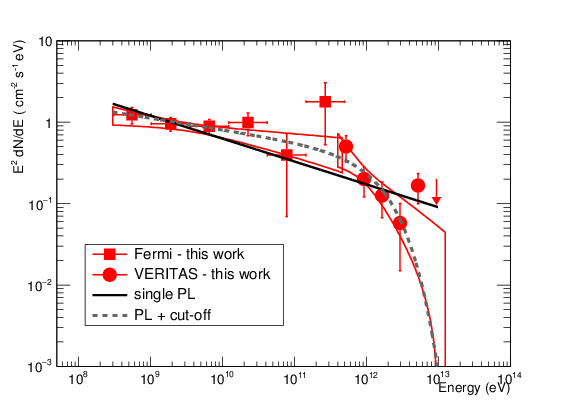}
        \caption{Fermi and VERITAS SEDs overlaid with combined fit results. Flux errors calculated from the error propagation of the fitting function are drawn as a $1 \sigma$ statistical error band in red lines.The Fermi-VERITAS combined fit results with a single power-law fit and a single power-law with a cut-off are shown with a black solid line and a gray short dashed line. }
   \label{Fig:Tycho_CombinedSpectrum_wFits}
 \end{figure}

The spectral index of 2.9 measured in the energy range of VERITAS (E$>$400 GeV) is somewhat softer than that measured in \textit{Fermi}'s energy range (E$<$500 GeV).  
This may indicate a cut-off of the gamma-ray spectrum around a few TeV or lower.  To test a possible spectral index change in the GeV--TeV gamma-ray range, we performed a goodness of fit test of the combined dataset of \textit{Fermi} and VERITAS with a single power-law $dN/dE = N_{0}(E/1\ \mathrm{TeV})^{-\Gamma}$ and a power-law with an exponential cut-off $dN/dE = N_{0}(E/1\ \mathrm{TeV})^{-\Gamma} \mathrm{e}^{-E/E_{cut}}$. Figure~\ref{Fig:Tycho_CombinedSpectrum_wFits} and Table~\ref{Tbl:Fermi_VERITAS_combinedFit} show the results. Both spectral forms are consistent within $2\sigma$, although we note that the statistical uncertainties are large. 

Recently, \cite{Morlino:2016kj} tried to explain the updated VERITAS spectrum by assuming that the gamma-ray emission is dominated by the dense northeastern region of the remnant. 
They took into account the presence of neutral hydrogen close to the shock, giving rise to Balmer-dominated shocks. They suggested that steep spectra such as those seen in Tycho occur due to the presence of charge-exchange reactions resulting from neutral hydrogen entering the shock front, and that a high neutral hydrogen fraction ($> 70$\%) could give spectra as steep as those seen in Tycho. This would primarily occur in dense regions where the neutral hydrogen fraction is highest and velocities lower. 
This model can provide low maximum energies, but these spectra would only arise in the denser regions, and not the remnant as a whole. 
\begin{table}[ht]
\caption{Fit results of the \textit{Fermi}-VERITAS combined dataset.}
\resizebox{\columnwidth}{!}{
\centering
\begin{tabular}{c c c}
\hline
 & Single & Single power- \\
 & power-law & law + cut-off \\ [0.5ex]
\hline
$N_{0}$\,($10^{-13}$\,cm$^{-2}$\,s$^{-1}$\,TeV$^{-1}$) & $1.72\pm0.29_{\mathrm{stat}}$ & $4.16\pm2.11_{\mathrm{stat}}$ \\
$\Gamma$ & $2.28\pm0.03_{\mathrm{stat}}$ & $2.14\pm0.08_{\mathrm{stat}}$ \\
$E_\mathrm{cut}$ (TeV) & N/A & $1.70\pm1.23_{\mathrm{stat}}$\\
$\chi^2$ / n.d.f. & 11.9 / 9 & 8.87 / 8 \\ [1ex]
\hline
\end{tabular}
}
\label{Tbl:Fermi_VERITAS_combinedFit}
\end{table}

Previous measurements from VERITAS reported a possible slight displacement of the emission toward the northeastern region of the remnant where a higher density of the surrounding medium was measured and a molecular cloud was observed along the line of sight. The updated centroid measurement is consistent with the previous result within $2\sigma$, but it is coincident with the center of the shell rather than offset toward the northeastern region. We also compared images of VERITAS data divided into two energy bins (one with energies lower than 800 GeV and the other with energies higher than 800 GeV), and found no significant centroid shift. 
Updated spatial distributions of gamma rays by \textit{Fermi} and VERITAS look similar and their centroids match within the $68\%$ confidence level.
Consequently, there is no statistical evidence that the GeV and TeV emission regions are different. 

We tested several spatial templates against the spatial distribution of the gamma rays observed by VERITAS. Table~\ref{Tbl:VERITAS_SpatialModelTest} shows the results of the likelihood ratio test. For each template, we smeared the source template with events drawn from gamma-ray simulations folded with the instrumental response of VERITAS. These simulated events were matched to the observational conditions for Tycho, namely the spectral index, elevation, and azimuthal angles.
A combined source and background model was then compared to a pure background model. 
As with the centroid study, only the data from the 2012--2015 seasons were used for this study because this data set has the highest statistics and the best angular resolution.

\begin{table}[ht]
\caption{Result of likelihood ratio tests of VERITAS spatial distribution with different spatial templates.}
\centering
\resizebox{\columnwidth}{!}{
\begin{tabular}{c c c}
\hline
Model & -$\Delta$ lnL & Significance ($\sigma$) \\ [0.5ex]
\hline
Point source & 15.5 & 5.6 \\
Uniform shell emission & 13.5 & 5.2 \\
NE quarter dominated emission & 12.3  & 5.0 \\ 
SW quarter dominated emission & 6.67  & 3.7 \\ 
Uniform molecular cloud emission & 0.804  & 1.3 \\ [1ex]
\hline
\end{tabular}
}
\label{Tbl:VERITAS_SpatialModelTest}
\end{table}

The geometric center of the X-ray emission observed by \textit{Chandra}, RA $0^h25^m19^s.9$ and declination $64^{\circ}8{\arcmin}18.2{\arcsec}$\citep{RuizLapuente:2004bh}, was used as the center of the remnant to define the model template. This location was also used as the location of the source to build a point source template. Although we do not expect to have a central compact object in Tycho, we added the point source template to test whether the current gamma-ray measurement can distinguish between point-like and extended source models for Tycho. To test for uniform shell emission, we used a simple ring shape template with uniform density with an inner radius of $0.063^{\circ}$ and an outer radius of $0.07^{\circ}$. We tested two additional templates: one in which only the northeastern (NE) quarter of the ring was included and another that used only the southwestern (SW) quarter of the ring. The case of NE quarter dominated emission was motivated to test for TeV emission originating only from the densest region of the shell. The SW quarter dominated emission's case was to test if the TeV emission would coincide with the brightest portion of the SNR at energies above 20 keV, as measured by \textit{NuSTAR}. For the uniform molecular cloud emission's case, we assumed the entire molecular cloud contour shown in Figure~\ref{Fig:Tycho_skymap_VERITAS} to be the source of the gamma rays.

The result from the point-source template shows the highest significance, followed by the uniform shell emission model. Meanwhile, the case for the uniform molecular cloud emission is the least favorable. To compare the models, we adopt a Bayesian approach, computing the posterior odds ratios under the assumption that the prior probability ratio between any two models is unity. We require a posterior odds ratio greater than 150 to provide ``decisive" evidence in favor of one model over another~\citep{Kass:1995eh}. 
We find no strong evidence to prefer the point source or the NE quarter dominated emission in comparison to the uniform shell emission. This can be caused by different or more complicated underlying gamma-ray distributions than the simple models considered here. It is difficult to draw strong conclusions, due to the limited statistics of the data sample.
When we compare the uniform shell emission model to the SW quarter dominated emission and the uniform molecular cloud emission, the posterior odds ratios are 960 and $3.4\times10^{5}$; thus we can rule out both models. 

\section{Conclusions}
We updated the high-energy gamma-ray studies of Tycho with a factor of two increased exposure for both VERITAS and \textit{Fermi}-LAT data. The improved low energy sensitivity of VERITAS allowed us to extend the TeV measurements toward lower energies. While the results are compatible with earlier measurements, we calculate a somewhat softer index compared to the previous measurement, which was calculated only for energies higher than 1 TeV. Both a single power-law and a single power-law with a cut-off describe the updated GeV--TeV fluxes consistently. The updated VERITAS result indicates a likely lower maximum particle energy than anticipated from theoretical studies developed to explain the previous data. These models may need to be revisited. The updated TeV centroid matches well with both the center of the remnant and the updated GeV centroid. The spatial distribution of the VERITAS source can be explained as either a point source or a uniform shell emission, while SW quarter dominated emission and uniform molecular cloud emission are disfavored. \\ 


\acknowledgements
This research is supported by grants from the U.S. Department of Energy Office of Science, the U.S. National Science Foundation and the Smithsonian Institution, and by NSERC in Canada. We acknowledge the excellent work of the technical support staff at the Fred Lawrence Whipple Observatory and at the collaborating institutions in the construction and operation of the instrument. The VERITAS Collaboration is grateful to Trevor Weekes for his seminal contributions and leadership in the field of VHE gamma-ray astrophysics, which made this study possible.
 
{\it Facility:} \facility{VERITAS}. \\

\bibliography{Tycho_2ndPaper}
\end{document}